\documentclass{PoS}
                        
\title{Computational efficiency of staggered Wilson fermions: A first look}

\ShortTitle{Computational efficiency of staggered Wilson fermions}

\author{\speaker{David H.~Adams}
\\
Division of Mathematical Sciences,        
Nanyang Technological University, Singapore 637371\\
        E-mail: \email{dhadams@ntu.edu.sg}}

\author{D\'{a}niel N\'{o}gr\'{a}di\\
Institute for Theoretical Physics, E\"{o}tv\"{o}s University,
H-1117 Budapest, Hungary\\
        E-mail: \email{nogradi@bodri.elte.hu}}

\author{Andrii Petrashyk
\thanks{Current address: Physics Dept., Columbia University, New York, USA}
\\
Division of Mathematical Sciences,        
Nanyang Technological University, Singapore 637371\\
        E-mail: \email{ap3115@columbia.edu}}

\author{Christian Zielinski
\\
Division of Mathematical Sciences,        
Nanyang Technological University, Singapore 637371\\
        E-mail: \email{zielinski@pmail.ntu.edu.sg}}

\abstract{

Results on the computational efficiency of 2-flavor staggered Wilson fermions compared to usual 
Wilson fermions in a quenched lattice QCD simulation on $16^3\times32$ lattice 
at $\beta=6$ are reported.  
We compare the cost 
of inverting the Dirac matrix on a source by the
conjugate gradient (CG) method for both of these fermion formulations, at the same pion masses, 
and without preconditioning.
We find that the number of CG iterations required for 
convergence, averaged over the ensemble, is less by a factor of almost 2 
for staggered Wilson fermions, with only a mild dependence on the pion mass.
We also compute the condition number of the fermion matrix and find that it is less by a factor
of 4 for staggered Wilson fermions.
The cost per CG iteration, dominated by the cost of matrix-vector multiplication for the 
Dirac matrix, is known from previous work to be less by a factor 2-3 for staggered Wilson
compared to usual Wilson fermions. Thus we conclude that staggered Wilson fermions are 4-6 times
cheaper for inverting the Dirac matrix on a source in the quenched backgrounds of our study.   
}

\FullConference{31st International Symposium on Lattice Field Theory - LATTICE 2013\\
		July 29 - August 3, 2013\\
		Mainz, Germany}

\begin{document}

\section{Introduction}

Staggered Wilson fermions are a novel lattice fermion formulation constructed by adding a 
``Wilson term'' to the staggered fermion action. 
Depending on the choice of this term, the number of fermion species (flavors)
is reduced from 4 to 2 or 1 \cite{DA(PLB),Hoel}.
A key idea in this construction is to choose the Wilson term in such a way that the
exact flavored chiral symmetry of staggered fermions becomes the {\em unflavored}
chiral symmetry of staggered Wilson fermions, broken explicitly by the staggered Wilson
term \cite{DA(PLB)}. The theoretical properties are then similar to usual Wilson fermions, 
but with a one component lattice fermion field just like staggered fermions. 

This development originated in the spectral flow approach to the staggered fermion index in 
\cite{DA(PRL)}, which led to a staggered version of the overlap Dirac operator describing 2 
fermion species. The staggered Wilson fermion arose as the kernel in that operator
\cite{DA(PLB)} (see also \cite{DA(proc10)} for further details). 
Another version of staggered Wilson fermions, describing 1 fermion species, was later 
proposed in \cite{Hoel}. 
The staggered Wilson terms in both cases are combinations of the flavored mass terms of
Golterman and Smit, which lift the degeneracy of the 4 staggered fermion flavors \cite{GS}. 

The main practical motivation for considering staggered Wilson fermions is that they are
potentially a more computationally efficient version of Wilson fermions, and can also be
used as kernel for constructing staggered versions of the overlap and domain wall 
fermions \cite{DA(PLB)} which should be cheaper than the expensive originals.
The breaking of chiral symmetry is less for staggered Wilson
fermions compared to usual Wilson fermions, as discussed later, which indicates that the improved
chiral properties of domain wall and overlap fermions should be cheaper to achieve
in the staggered versions.  
Also, as new lattice fermion formulations, the staggered versions of Wilson, domain wall 
and overlap fermions can be used for testing universality in lattice QCD.

Here we report on a first test of the computational efficiency of staggered Wilson fermions.
We compare the cost (computation time) of inverting the Dirac matrix on a source for 2-flavor staggered 
Wilson and usual Wilson fermions in quenched lattice QCD with fixed physical
volume and lattice spacing.
The ratio of these computation times is our efficiency measure. 
To be meaningful, the comparison should be done at fixed values of a physical quantity. 
We take this to be the pion mass. I.e. our cost comparison for inverting the Dirac matrix
is done with the staggered Wilson and usual Wilson fermions having different bare quark masses 
such that the pion masses are the same. Thus our efficiency measure is a function of the pion mass.    
 
In the following we first discuss the theoretical background for staggered Wilson
fermions, and then present the results of our numerical study. We conclude with remarks on 
prospects for the computational efficiency of staggered overlap and domain wall fermions. 
In particular we discuss implications of our results for the computational efficiency
of staggered overlap fermions. The results give hope that the efficiency speed-up 
in thermalized backgrounds may be significantly greater on lattices of our size or larger 
compared to the modest speed-up by a factor 2-3 found in the numerical study of 
De Forcrand {\em et al.} on a smaller lattice \cite{Forcrand}. 


\section{Theoretical Background}

At first sight, introducing a staggered Wilson term appears to be problematic since it breaks
some of the staggered fermion symmetries. The concern in this situation is that
new counterterms can arise which satisfy the remaining symmetries; these would then need to be 
included in the bare action from the beginning and fine-tuned to reproduce continuum QCD.

For the 2-flavor staggered Wilson fermion the situation turns out to be fortuitous: it breaks the 
`shift' symmetries of the staggered fermion (corresponding to certain flavor symmetries in the
continuum), but only one new counterterm of mass-dimension $\le4$ arises, and its
effect on the 2 physical fermion species is simply a wavefunction renormalization 
\cite{DA(PLB)}. Therefore it does not need to be included in the bare action, and no fine-tuning
is required (besides the usual fine-tuning of the usual mass term that one also does for usual 
Wilson fermions).   

However, the situation is worse for the 1-flavor staggered Wilson fermion of \cite{Hoel}.
Besides breaking the shift symmetries, it also breaks lattice rotation symmetry. A subgroup of the
latter survives, but it is not enough to prevent a new gluonic counterterm of mass-dimension
4 from arising \cite{DA(prep)}. This term needs to be included in the bare action
and fine-tuned, thus reducing the attractiveness of the 1-flavor staggered Wilson fermion for practical 
use.
 
Besides the issue of broken symmetries, there is also at first sight a chirality problem for
staggered Wilson fermions: The unflavored staggered version of $\gamma_5$ violates the property 
$\gamma_5^2=1$ by $O(a)$ effects, and also one does not have the $\gamma_5$-hermiticity property 
$\,\gamma_5 D_W\gamma_5=D_W^{\dagger}\,$ of the usual Wilson Dirac operator $D_W$. The solution
is to use the {\em flavored} $\gamma_5$ that gives the exact flavored chiral symmetry of staggered 
fermions as the {\em unflavored} $\gamma_5$ of the staggered Wilson fermion \cite{DA(PLB)}. 
This is possible because the flavored $\gamma_5$ acts in an unflavored way on the physical fermion 
species of the staggered Wilson fermion in both the 2- and 1-flavor cases.   
The staggered Wilson Dirac operators are then $\gamma_5$-hermitian and have the other 
Wilson-like properties required to construct staggered versions of domain wall and overlap fermions 
\cite{DA(PLB)}.

A significant drawback of the 2-flavor staggered Wilson formulation is that the SU(2) vector
and chiral symmetries of 2 flavors of usual Wilson fermions are broken
by lattice effects, just like the SU(4) symmetries of the staggered fermion. However, the 
unbroken symmetries, which include all the flavored rotation symmetries, are still enough to
impose some of the same consequences as the SU(2) symmetries. E.g. they are enough 
to ensure a degenerate triplet of pions \cite{Sharpe(Kyoto-talk)}. 

On the other hand, regarding {\em flavor-singlet} chiral symmetry,
2-flavor staggered Wilson fermions have no disadvantage compared to usual Wilson fermions. 
Thus, for flavor-singlet physics, and in particular the challenge of high-precision computation of 
the $\eta'$ mass, staggered Wilson fermions and the associated staggered versions of domain wall 
and overlap fermions offer increased computational efficiency with no significant theoretical 
drawbacks as a lattice formulation for the light $u$ and $d$ quarks.
This also appears to be the case for their use to calculate bulk quantities in QCD thermodynamics. 
So these are at least two important arenas where we envisage that staggered Wilson-based fermions 
will be advantageous compared to usual Wilson-based fermions. For other challenges where flavored 
chiral symmetry plays a more important role, such as in the computation of hadronic matrix 
elements in weak interaction processes, it remains to be seen whether or not staggered 
Wilson-based fermions are advantageous compared to currently used lattice fermions. This will 
depend both on how computationally efficient the staggered Wilson-based fermions are, and how 
severe the consequences of their SU(2) flavor symmetry breaking turns out to be.

We omit the explicit expression for the 2-flavor staggered Wilson Dirac operator here. It can be 
found in \cite{DA(PLB)}. A detailed treatment of the theoretical aspects discussed here,
and related aspects, is currently in preparation \cite{DA(prep)}.

\section{Numerical Results}

Our quenched simulations were done on the $16^3\times32$ lattice with 200 configurations
generated at $\beta=6$,
using the Chroma/QDP software for lattice QCD \cite{chroma} (we hacked the asqtad staggered
fermion code there to make the code for staggered Wilson fermions).

Our results for the pion mass as a function of the bare quark mass are shown in Fig.~1(a).
A check on the validity of the staggered Wilson formulation is that it 
exhibits a linear relation $m_{\pi}^2\sim m$ in accordance with Chiral Perturbation Theory.
Significantly, the  additive mass renormalization is seen to be less for staggered Wilson 
compared to usual Wilson fermions.
Similar results for the pion mass were earlier obtained independently by us and by De Forcrand
{\em et al.} \cite{Forcrand,DA(Kyoto-talk)}. Our ``pion'' operator in the staggered Wilson
case is the same as the one described in \cite{Forcrand}.

Knowing the relations between pion mass and the bare quark mass, we can proceed to measure
the costs of inverting the staggered Wilson and usual Wilson Dirac matrices on a source $\chi$ 
at the same pion mass. This was done by solving 
\begin{equation}
(D+m)^{\dagger}(D+m)\psi=(D+m)^{\dagger}\chi^{(\alpha)}.
\label{2}
\end{equation}
using the conjugate gradient (CG) method, without preconditioning\footnote{Preconditioning 
speeds up the computation for Wilson fermions by more than a factor 2. It can also be done 
for staggered Wilson fermions, but we did not implement it yet, so the cost comparison here 
is without preconditioning.}, with $\chi^{(\alpha)}$ being the point source at the origin with 
(spin and) color component $\alpha$. We averaged the cost over the different $\alpha$'s 
(3 sources for staggered Wilson vs 12 for usual Wilson). Hence our results also give a cost
comparison for computing the fermion propagator matrix for staggered Wilson and usual Wilson
fermions.  

The cost (computation time) ratio decomposes as
\begin{equation}
\frac{\mbox{cost($W$)}}{\mbox{cost($SW$)}}
=\Big(\frac{\mbox{iters($W$)}}{\mbox{iters($SW$)}}\Big)
\times\Big(\frac{\mbox{cost per iter($W$)}}{\mbox{cost per iter($SW$)}}\Big)
\label{3}
\end{equation}
where `$W$' and '$SW$' refers to Wilson and staggered Wilson, respectively, and 'iters' is the 
number of CG iterations for convergence with a given CG residual $\epsilon$. When considering
the fermion propagator, an extra factor 4 should be included on the right-hand side
because of the 12 vs 3 sources for the usual Wilson vs staggered Wilson case.
But this should not be included when using (\ref{3}) as an estimate of the cost ratio for 
generating configurations in dynamical fermion simulations of lattice QCD, which is the main 
quantity of interest for ascertaining the computational efficiency of staggered Wilson fermions.  

Our results for the CG iterations ratio, i.e. the first ratio on the right-hand side of (\ref{3}),
averaged over the configurations of our ensemble, are shown in Fig.~1(b) for a selection
of our smaller pion masses.
The number of CG iterations required with staggered Wilson fermions
is seen to be less by almost a factor 2 compared to usual Wilson fermions. Furthermore, the 
iterations ratio has only a mild dependence on the pion mass, even though the number of CG 
iterations in the staggered Wilson and usual Wilson cases both depend strongly on it. 
The results for the two lowest pion masses are less reliable --
we noticed some instability in the CG computations for these cases, affecting the number
of iterations required for convergence.

A related quantity of interest is the ratio $\kappa_W^{1/2}/\kappa_{SW}^{1/2}$, where
$\kappa=\lambda_{max}/\lambda_{min}$ is the condition number of the fermion matrix
$(D+m)^{\dagger}(D+m)$ in (\ref{2}). The $\kappa^{1/2}$ ratio gives an estimate of the CG iterations
ratio, which is expected to be better when the CG residual $\epsilon$ is smaller. 
(See, e.g., \cite{painlessCG}.) We computed the condition number as a function of the pion mass
in both the staggered Wilson and usual Wilson cases, and our results for the $\kappa^{1/2}$
ratio are also shown in Fig.~1(b). They are seen to agree approximately with the CG iterations 
ratios, with better agreement for smaller $\epsilon$ as expected, except for the smaller pion 
masses where the above-mentioned CG instabilities occur. 


The $\kappa^{1/2}$ ratio $\approx2$ in Fig.~1(b) can be compared with the free field case
with same bare mass $m$ for both formulations:
$\kappa_{W,free}^{1/2}/\kappa_{SW,free}^{1/2}\,\stackrel{m\to0}{=}\,8/\sqrt{5}\,\approx\,3.58$.

\begin{figure}
\centerline{ 
\includegraphics[width=0.39\textwidth]{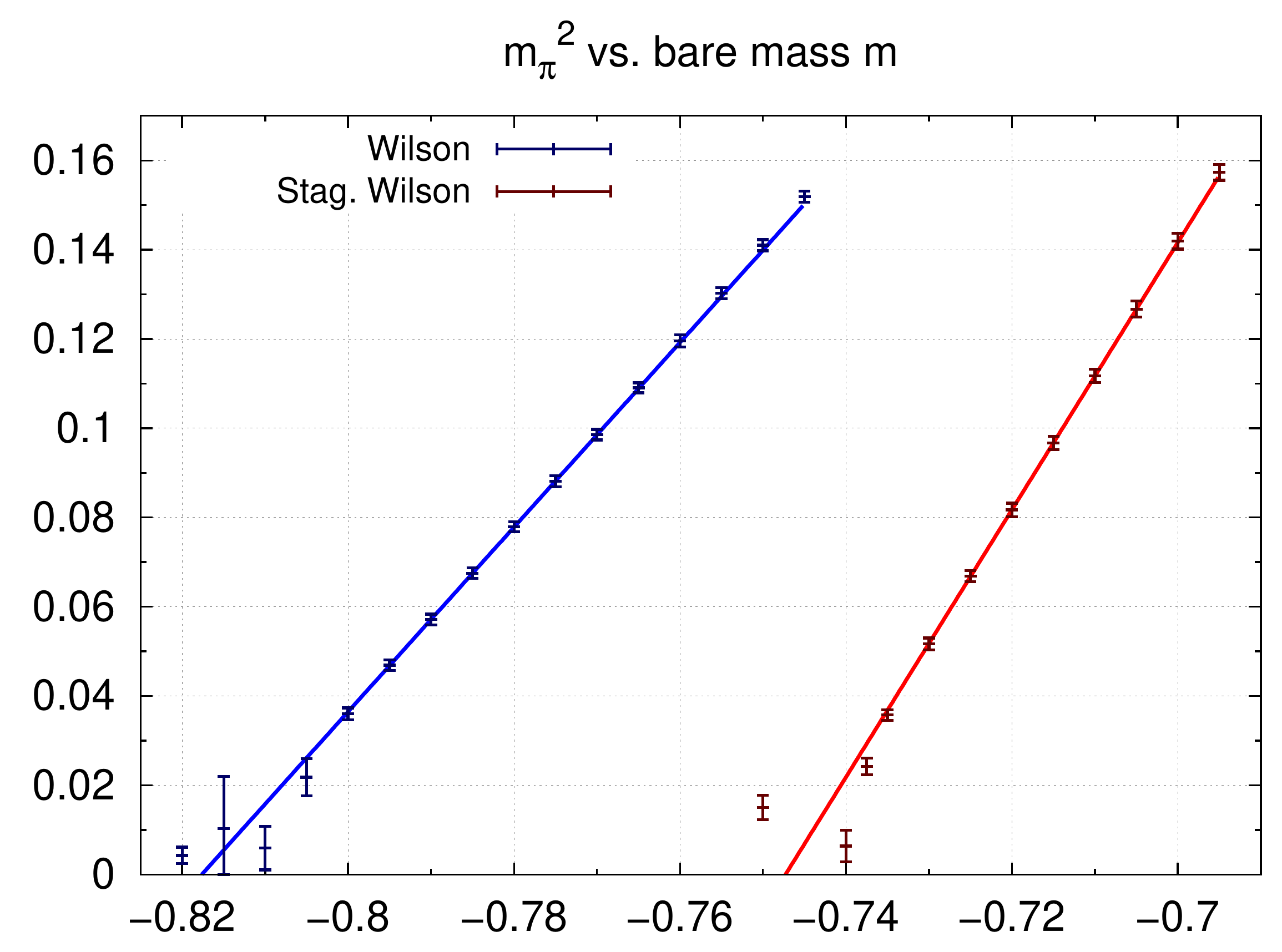} \hfill 
\includegraphics[width=0.39\textwidth]{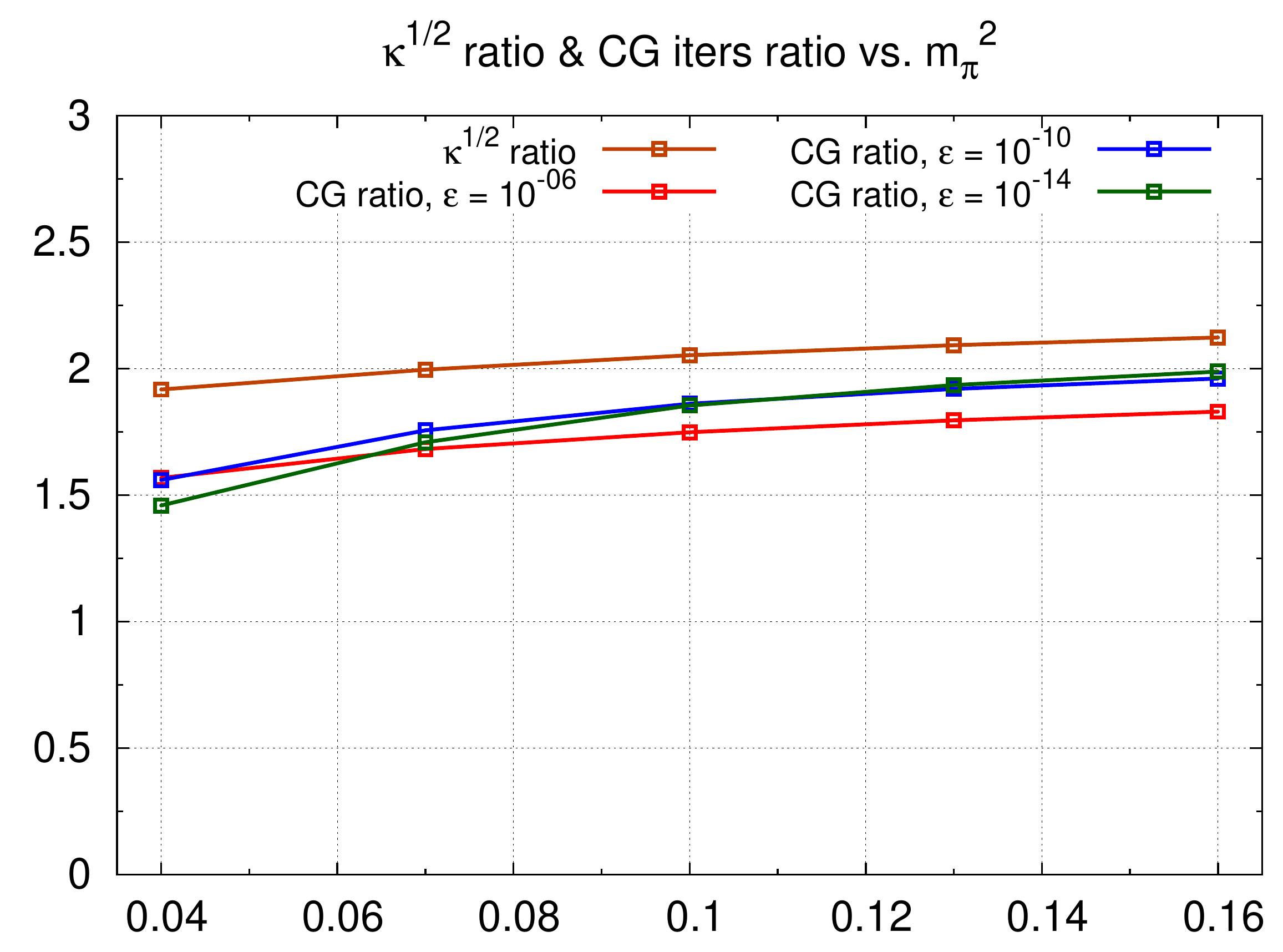} \hfill 
\includegraphics[width=0.39\textwidth]{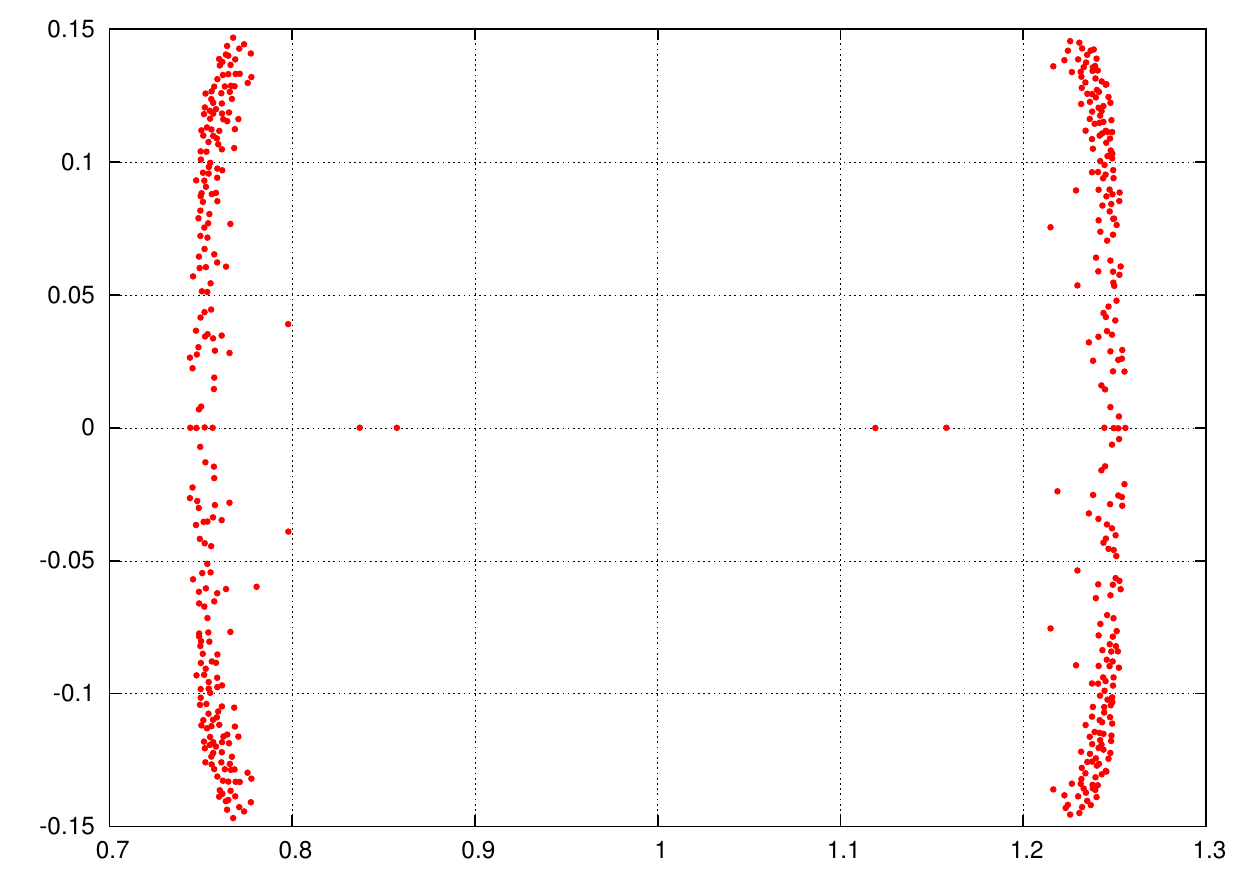}
}
\caption{(a) Left: $m_{\pi}^2$ as a function of the bare quark mass $m$ in lattice units.
Straight lines fitted through the data points with $0.05<m_{\pi}^2<0.1$. 
(b) Middle: averaged $\kappa^{1/2}$ ratio and CG iterations ratios as functions of $m_{\pi}^2$.
Data joined by lines to guide the eye.
(c) Right: The low-lying staggered Wilson eigenvalue spectrum in one of our 
backgrounds.}
\end{figure}

Regarding the second ratio in (\ref{3}), the cost per CG iteration is dominated by the cost
of matrix-vector multiplication with the fermion matrix $(D+m)^{\dagger}(D+m)$. It can be estimated 
as being proportional to the number of floating point operations required (this is only a rough 
first estimate though, since it does not take memory bandwidth into account).
Consequently, for the ratio we have the estimate 
\begin{equation}
\frac{\mbox{cost per iter($W$)}}{\mbox{cost per iter($SW$)}}\;\approx\;
4\times\frac{\mbox{flops($W$)}}{\mbox{flops($SW$)}}=4\times\frac{1392}{1743}\;\approx\;3.2.
\label{4}
\end{equation} 
Here `flops' denotes the number of floating point operations per lattice site for matrix-vector
multiplication with the lattice Dirac matrix $D$. We have plugged in the known value 1392
for the usual Wilson case, and the value 1743 that we find for the
staggered Wilson $D$. The factor 4 after the equality in (\ref{4}) is because the usual Wilson 
Dirac matrix is 4 times larger than the staggered Wilson one.

These numbers can be roughly understood as follows \cite{Forcrand}: The staggered Wilson action
couples each lattice site to $8+16$ other sites, whereas in the usual Wilson case it is 
$8\times2$ (the factor 2 is for the 2 Dirac spin components after the spin projections). 
Including the above-mentioned factor 4, we then get the ratio estimate
$4\times(8\times2)/(8+16)\approx2.67$ \cite{Forcrand}. 
This is slightly smaller than in (\ref{4}) because the small but non-negligible cost 
of spin decomposition and reconstruction in the spin projection trick has not taken into account. 

A speed-up factor of order 2 for the matrix-vector multiplication in the staggered Wilson case
was explicitly found in the numerical study of \cite{Forcrand}. In light of this and the
estimate (\ref{4}), we assume the achievable speed-up factor to be 2-3. (We are far from achieving this
speed-up in our own study, but that is no doubt due to shortcomings in our implementation
of staggered Wilson fermions in Chroma. We hope to improve it in future.)
Combining this with the speed-up factor $\approx2$ for the number of CG iterations, we get 
an estimated speed-up factor of 4-6 for inverting the Dirac matrix on a source in the
staggered Wilson case.

Finally, in Fig.~1(c) we show the low-lying 
spectrum of the staggered Wilson Dirac operator in one of the gauge field backgrounds of our 
ensemble. The situation is clearly better than in the small 
$8^4$ lattice results of \cite{Forcrand}.
E.g. the separation between the physical and doubler branches is $\sim0.5$ in our case
compared to $\sim0.3$ in \cite{Forcrand}. (See the 2nd panel in Fig.~5 of the first article
in \cite{Forcrand}.)

\section{Conclusions}

The estimated speed-up factor of 4-6 in the staggered Wilson case for inverting the Dirac matrix 
on a source gives tentative encouragement for the prospects of cheaper lattice QCD simulations 
with dynamical staggered Wilson fermions. 
However, it should be remembered that this is only a quenched 
exploratory study. It should be followed up
in future by systematic studies of the computational efficiency in full QCD simulations, 
with $O(a)$ improvement of staggered Wilson fermions and using the HISQ version of the usual
staggered part of the action to reduce $O(a^2)$ effects, and with 
smeared links, so that a realistic comparison with presently used improved Wilson fermions
can be made. $O(a)$ improvement for staggered Wilson fermions via a version of the clover term 
has been done and will be reported elsewhere \cite{DA(prep)}. 
A different proposal for $O(a)$ improvement was recently studied, along with 
smearing, in \cite{Durr}. 

On the other hand, our results give a complete solution to the problem of estimating the
cost ratio for computing the fermion propagator in our quenched study. An extra factor 4
should be included in (\ref{3}) for this, so we get an estimated speed-up factor of between
16-24 for computing the fermion propagator with staggered Wilson fermions.  

The results here also have positive implications for the potential efficiency
of staggered versions of domain wall and overlap fermions. They add to the evidence 
that staggered Wilson fermions are more chiral than usual Wilson fermions, as discussed 
below. The cost of overlap fermions is expected to be reduced when a more chiral kernel
is used \cite{Biet}, and in the case of staggered fermions one can hope to achieve the 
same level of approximate chirality with a smaller 5th dimensional lattice size.

The more chiral nature of staggered Wilson fermions is
strikingly clear in the free field case where the spectrum is 
close to the Ginsparg-Wilson (GW) circle \cite{Forcrand}. In the interacting case, spectrum
computations in thermalized backgrounds on a small $8^4$ lattice found that the spectrum 
collapses away from the GW circle into a vertical strip, with the physical branch becoming 
diffuse and approaching the doubler branch \cite{Forcrand}. This was attributed to large 
fluctuations in the 4-link staggered Wilson term. However, our present results indicate that
the chiral GW-like nature of staggered Wilson fermions does persist in the interacting case
on larger lattices. If it did not, the condition number of the fermion matrix would be expected 
to be larger in the staggered Wilson case, due to the diffuse spectrum giving rise to near zero 
eigenvalues when the pion mass is small. 
But we found here that the condition number is smaller by a factor 4 for staggered Wilson fermions,
cf.~Fig.~1(b). The better-behaved spectrum in Fig.~1(c), and the smaller additive mass 
renormalization for staggered Wilson fermions in Fig.~1(a), is further
evidence that the chiral GW-like nature persists in the thermalized backgrounds of our study.

The preceding has implications when assessing the potential of reduced cost for staggered 
overlap fermions. The computational cost of staggered overlap fermions vs usual overlap
fermions for inverting the Dirac matrix on a source was investigated on a 
$12^4$ lattice in \cite{Forcrand}. In the free field case, a large speed-up factor of almost 10
was found, confirming the expectations discussed above for reduced cost of overlap fermions
when a more chiral kernel is used.
However, in thermalized $\beta=6$ backgrounds the speed-up factor was found to be dramatically
reduced to 2-3 \cite{Forcrand}. This was attributed to the aforementioned ruining of the GW-like 
nature of the staggered Wilson kernel in the interacting case. But in the present case,
on our larger lattice, our results indicate, as discussed above, that the GW-like nature
survives to a reasonable extent. Thus one can hope for 
much better efficiency of staggered overlap fermions on our lattice and larger lattices
in the interacting case. In this way our results suggest that the modest speed-up factor 
2-3 found in \cite{Forcrand} is not indicative of the true potential of reduced cost
with staggered overlap fermions.

  
We hope to clarify the situation in the near future by
repeating the present study for staggered overlap fermions (on our larger $16^3\times32$ lattice), 
so as to determine the cost ratio as a function of the pion mass. 
Similar studies for staggered domain wall fermions are also planned.

 
{\bf Acknowledgments.}
D.A. and A.P. thank the Yukawa Institute, Kyoto, for hospitality and support 
at the workshop ``New Types of Fermions on the Lattice", which spurred on this work.
D.A. thanks Philippe de Forcrand for feedback on a previous version of the paper.
D.A. is supported by AcRF grant RG61/10. 
D.N. is supported by the EU under grant (FP7/2007-2013)/ERC No
208740 and by OTKA under grant OTKA-NF-104034.


\begin{thebibliography}{99}
\bibitem{DA(PLB)} 
D.H.~Adams, Phys.~Lett.~B 699 (2011) 394 [arXiv:1008.2833]

\bibitem{Hoel} 
C.~Hoelbling, Phys.~Lett.~B 696 (2011) 422 [arXiv:1009.5362]

\bibitem{DA(PRL)}
D.H.~Adams, Phys.~Rev.~Lett.~104:141602 (2010) [arXiv:0912.2850]

\bibitem{DA(proc10)}
D.H.~Adams, PoS LATTICE2010 (2010) 073 [arXiv:1103.6191]

\bibitem{GS}
M.F.L.~Golterman and J.~Smit, Nucl.~Phys.~B 245 (1984) 61

\bibitem{Forcrand}
Ph.~de Forcrand, A.~Kurkela and M.~Panero, PoS LATTICE2010 (2010) 080 [arXiv:1102.1000];
JHEP 1204 (2012) 142 [arXiv:1202.1867]

\bibitem{DA(prep)}
D.H.~Adams, unpublished work (articles in preparation).

\bibitem{Sharpe(Kyoto-talk)}
S.~Sharpe, talk at Yukawa Institute Workshop ``New Types of Fermions 
on the Lattice'', Kyoto 2012. 

\bibitem{chroma}
R.G.~Edwards and B.~Joo, 
Nucl.~Phys.~Proc.~Suppl.~ 140 (2005) 832 [hep-lat/0409003]

\bibitem{DA(Kyoto-talk)}
D.~Adams, talk at Yukawa Institute Workshop ``New Types of Fermions 
on the Lattice'', Kyoto 2012. 

\bibitem{painlessCG}
J.R.~Shewchuk, ``An Introduction to the Conjugate Gradient Method Without Agonizing Pain.''
http://www.cs.cmu.edu/\textasciitilde{}quake-papers/painless-conjugate-gradient.pdf

\bibitem{Durr}
S.~Durr, Phys.~Rev.~D87 (2013) 114501 [arXiv:1302.0773]

\bibitem{Biet}
W.~Bietenholz, Eur.~Phys.~J.~C6 (1999) 537 [hep-lat/9803023]







\end{thebibliography}
\end{document}